\documentclass[12pt]{article}
\usepackage{bbm}
\usepackage{bbding}
\usepackage{amsmath}
\usepackage{amssymb}
\usepackage{amsfonts}
\usepackage{mathrsfs}
\usepackage{mathrsfs, amssymb}
\usepackage{booktabs}
\usepackage[perpage,symbol*]{footmisc}
\usepackage{graphicx}
\usepackage{multirow}
\usepackage[level]{datetime}
\usepackage[a4paper]{geometry}

\newcounter{theorem}

\newcounter{lemma}

\newcounter{remark}

\newcounter{example}

\newcounter{definition}

\newcounter{corollary}

\newcounter{proposition}

\newcounter{assumption}

\newcounter{condition}

\newcounter{algorithm}

\makeatletter
\newcommand*{\indep}{%
  \mathbin{%
    \mathpalette{\@indep}{}%
  }%
}
\newcommand*{\nindep}{%
  \mathbin{
    \mathpalette{\@indep}{\not}
  }%
}
\newcommand*{\@indep}[2]{%
  \sbox0{$#1\perp\m@th$}
  \sbox2{$#1=$}
  \sbox4{$#1\vcenter{}$}
  \rlap{\copy0}
  \dimen@=\dimexpr\ht2-\ht4-.2pt\relax
  \kern\dimen@
  {#2}%
  \kern\dimen@
  \copy0 
}
\makeatother

\begin{document}
\baselineskip 18pt
\begin{center}
{\large\bf Transfer Learning for High Dimensional Robust Regression}
\end{center}

\begin{center}
{ \textbf{\small{ Xiaohui Yuan$^*$}}}\\
{ \small{School of Mathematics and Statistics, Changchun University of Technology, China}}\\
{ \small{\texttt{\textsc{yuanxh@ccut.edu.cn}}}}

{ \textbf{\small{ Shujie Ren}}}\\
{ \small{School of Mathematics and Statistics, Changchun University of Technology, China}}\\
{ \small{\texttt{\textsc{renshujie0929@163.com}}}}
\end{center}
\begin{center}
This version: \usdate\today
\end{center}

\footnotetext{$^*$Corresponding author, $^\dag$ equal authors contribution.}

\begin{abstract}
{\small Transfer learning has become an essential technique for utilizing information from source datasets to improve the performance of the target task. However, in the context of high-dimensional data, heterogeneity arises due to heteroscedastic variance or inhomogeneous covariate effects. To solve this problem, this paper proposes a robust transfer learning based on the Huber regression, specifically designed for scenarios where the transferable source data set is known. This method effectively mitigates the impact of data heteroscedasticity, leading to improvements in estimation and prediction accuracy. Moreover, when the transferable source data set is unknown, the paper introduces an efficient detection algorithm to identify informative sources. The effectiveness of the proposed method is proved through numerical simulation and empirical analysis using superconductor data.
}
\end{abstract}

\section {Introduction} \label{sec1} \setcounter {equation}{0}
\def\theequation{\thesection.\arabic{equation}}

With the rapid advancement of science and technology, machine learning, as one of the important topics of artificial intelligence, simulates the learning process of human by using computers and has become more and more popular. It needs ample training data to effectively accomplish various tasks. However, acquiring such data can be challenging, particularly in specialized domains. For example, in the agricultural fields, it's necessary to collect sufficient data on soil quality, climatic conditions, and crop growth, but this data might not be readily available in a timely manner. In most cases, there might be some relevant real datasets accessible alongside the limited data for the target task. Transfer learning offers a powerful technique by harnessing external data to boost its performance. It has been widely used in various aspects, including computer vision \cite{Kulis2011}, natural language processing \cite{Ruder2019}, agriculture \cite{Zhao2022}, and medicine \cite{Shojaie2022, Wang2022}, among others.

Nowadays, the theoretical research of model-based transfer learning has made great progress. For instance, linear regression model \cite{Li2022}, quantile regression model \cite{Huang2022, Zhang2022}, generalized linear models \cite{Tian2023}, composite quantile regression model \cite{Li2024}. In these models, transfer learning efficiently utilizes knowledge from the source data to guide the learning task of the target data, thus improving the efficiency of data utilization and providing valuable information and inspiration for the target domain. However, besides above models, there is a class of robust regression models that have not been studied in the field of transfer learning. Therefore, we study the transfer learning scenarios based on high-dimensional robust regression model.

In complex real-world data, heterogeneity often exist. Huber \cite{Huber1964} proposes a Huber loss function firstly and proves the asymptotic robustness of the proposed estimation. This loss function not only reduces the influence of outliers but also alleviates their impact on the regression model. As a robust regression method, Huber regression is specifically designed to minimize the effects of outliers, making it a compelling choice for handling these data. Extensive scholarly researchers support the efficacy of Huber regression, particularly in data with outliers. To handle big data with outliers or covariates that contain heavy tail distributions,  Sun et al. \cite{Sun2020} propose an adaptive Huber regression to obtain the robust estimation of parameter. To address challenges in regression on large data streams and outlier management, Tao \& Wang \cite{Tao2022} propose an online updating Huber robust regression algorithm. Additionally, in order to solve the problem of data privacy, Luo et al. \cite{Luo2022} introduce a robust distributed Huber regression to handle distributed data.

For high-dimensional Huber regression, a regularization process is often necessary. Yi \& Huang \cite{Yi2017} propose a semismooth newton coordinate descent (SNCD) algorithm designed for robust regression with Huber loss and quantile under elastic-net penalization, this method works well in simulations and real data and establishes the convergence properties of this algorithm. Moreover, Pan et al. \cite{Pan2021} propose a iteratively reweighted $\ell_{1}$-penalized adaptive Huber regression. Liu et al. \cite{Liu2024} propose a robust regression approach to analyze high-dimensional imaging data. This paper presents a robust transfer learning method based on Huber regression for scenarios where the transferable source dataset is known in high-dimensional data. Furthermore, when the transferable source data set is unknown, an effective source detection algorithm is introduced to identify suitable transferable source datasets. Simulation and experiments with real data validate the effectiveness of the proposed approach.

The rest of the paper is organized as follows. In Section 2, we introduce the Huber regression model and establish a robust transfer learning based on Huber regression model. In the scenario of the unknown transferable source datasets, we employ an efficient method to detect the transferable source data. In Section 3, We conduct several numerical simulations, which verify the proposed method for homogeneous and heterogeneous data, respectively. In Section 4, the effectiveness of the proposed method is validated through a real data. Section 5 presents a comprehensive summary of the whole article.

\section{Methodology} \setcounter {equation}{0}
\def\theequation{\thesection.\arabic{equation}}

\subsection{Huber regression model}

Considering a linear regression model:
\begin{eqnarray}
y_i=\beta_{0}+x^T_i\beta_{1}+\epsilon_{i}=z^T_i\beta+\epsilon_{i},\ i=1,\ldots,n
\end{eqnarray}
where $y_i$ is a response variable, $x_i$ is a $p$ dimensional covariate and $z_i=(1,x^T_i)$, $\beta=(\beta_{0},\beta^T_{1})^T$ is a $p+1$ vector of unknown parameters, and $\epsilon_{i}$ is an error term. In statistics, the ordinary least square (OLS) method is generally used to solve the linear regression model, but it assumes a normal distribution of errors and is sensitive to outliers. In contrast, Huber regression does not need to assume a specific distribution for the error terms, which helps effectively reduce the impact of outliers on parameter estimation. The Huber loss function is defined as
\begin{eqnarray}
\ell(t) = \begin{cases}
    |t| - 0.5\gamma & \text{if } |t| > \gamma, \\
    0.5{t^2}/{\gamma} & \text{if } |t| \leq \gamma.
\end{cases}
\end{eqnarray}
\\
where $\gamma$ is threshold parameter. This loss allows us to combine analytical tractability of the squared loss used in OLS regression and robustness of the absolute loss in least absolute deviations (LAD) regression.

Moreover, Huber regression model exhibits greater robustness in handling high-dimensional data containing outliers. According to Yi and Huang \cite{Yi2017}, we can obtain the estimation of parameter by solving the convex optimization problem
\begin{eqnarray}
\min_\beta\frac{1}{n}\sum_{i}\ell(y_i-z^T_i\beta)+{\lambda}P_{\alpha}(\beta),
\end{eqnarray}
where $\lambda$ is a penalty parameter, and $P_{\alpha}(\cdot)$ is the elastic-net penalty $P_{\alpha}(\beta)=\alpha{\|\beta\|_{1}}+0.5(1-\alpha){\|\beta\|^2_{2}}$.

To achieve a robust and efficient estimator for solving the parameters of the Huber regression model, we employ the SNCD algorithm developed by Yi \& Huang \cite{Yi2017}. The SNCD algorithm, integrated into the publicly available hqreg package (available at https://cran.r-project.org/web/packages/hqreg/index.html), offers several advantages. It combines semismooth newton with coordinate descent algorithms, optimizing computational efficiency. Specifically designed for high-dimensional data with heavy-tailed errors, SNCD is known to converge effectively under certain conditions. In this study, we utilize the SNCD algorithm to address the optimization problem (2.3).

\subsection{Transfer learning based on Huber regression model}
\def\theequation{\thesection.\arabic{equation}}

This paper explores a robust transfer learning approach utilizing the Huber regression model. Suppose that the target dataset $\{(z^{(0)}_{i},y^{(0)}_{i})\}^{n_{0}}_{i=1}$ is independent and identically distributed ($i.i.d.$) according to the target population $(Z^{(0)},Y^{(0)})$. And the source datasets $\{(z^{(k)}_{i},y^{(k)}_{i})\}^{n_{k}}_{i=1}$, $k\in \{1,\ldots,S\}$, are $i.i.d.$ copies of the source population $(Z^{(k)},Y^{(k)}),k\in \{1,\ldots,S\}$, where $S$ is the number of source datasets. For $k\in \{0,\ldots,S\}$, we consider the Huber regression model:
\begin{eqnarray}
y^{(k)}_{i}=z^{(k)T}_{i}\beta^{(k)}+\epsilon^{(k)}_{i}, \ i=1,\ldots,n_{k},
\end{eqnarray}
where $\beta^{(k)}$ is the $p+1$ vector of unknown regression coefficients based on the $k$-th dataset.

The main goal is to fit a better target model by extracting useful information from source datasets to assist target dataset. Assume the target model is $\ell_{0}$-sparse, which satisfies $\|\beta^{(0)}\|_{0}=s\ll {p+1}$, indicating that only $s$ out of $p+1$ variables significantly influence the target response. Intuitively, if $\beta^{(k)}, k\in \{1,\ldots,S\}$ is close to $\beta^{(0)}$, the $k$-th source dataset will be transferred. The tool to measure the similarity between $\beta^{(k)}$ and $\beta^{(0)}$ is the distance between two vectors. The concept of distance can be approached in various ways, such as $\ell_{1}$ norm, $\ell_{2}$ norm, cosine value, sine value and so on. The most common is $\ell_{1}$ norm and this paper use it to measure similarity between two vectors and expressed as $||\beta^{(k)}-\beta^{(0)}||_{1}$. Denote $\bigtriangleup^{(k)}=\beta^{(k)}-\beta^{(0)}$, if $||\bigtriangleup^{(k)}||_{1}\leq{h}$, $h$ is a non-negative number, the k-th source data set will be transferred. The indicator sets from all source datasets that can be transferred are denoted by $\Lambda_{h}=\{k:||\bigtriangleup^{(k)}||_{1}\leq{h},1{\leq}k{\leq}S\}$, and their cardinality is represented by $|\Lambda_{h}|$. The smaller the value of $h$, the more efficient the correspondent transferable source set becomes. Meanwhile, the number of $\Lambda_{h}$ will reduce. Therefore, we need to choose a proper $h$ to strike a balance between $h$ and $\Lambda_{h}$.

\subsubsection{Two-step transfer learning}
\def\theequation{\thesection.\arabic{equation}}

We introduce a two-step transfer learning algorithm when $\Lambda_{h}$ is known, called Oracle. The case that $\Lambda_{h}$ is unknown will be discussed in the next section. Figure 1 shows a schematic of two-step transfer learning.

Motivated by Li and Song \cite{Li2024}, a two-step transfer learning algorithm for Huber regression model is proposed. The main idea is to fit a Huber regression model by imposing elastic-net penalty on Huber loss based on $\Lambda_{h}$ and the target dataset, then correct the bias on the target dataset by imposing elastic-net penalty. The oracle algorithm is presented in Algorithm 1. Specifically, we firstly develop a rough estimator for elastic-net-penalized Huber regression, which can be defined as
\begin{eqnarray}
\hat{w}_{\Lambda_{h}}=\arg\min_{w\in {R^p}}\left\{\frac{1}{n_{\Lambda_{h}}+n_{0}}\sum_{k\in \Lambda_{h}}\left(\sum^{n_{k}+n_{0}}_{i=1}\ell(y_i-z^T_iw)\right)+\lambda_{w}P_{\alpha}(w)\right\},
\end{eqnarray}
where $\lambda_{w}$ is the tuning parameter and $n_{\Lambda_{h}}=\sum_{k\in \Lambda_{h}} n_{k}$. Although $\hat{w}_{\Lambda_{h}}$ combines the transferable source datasets with the target data, the $\lambda_{w}$ is a crude estimator. Secondly, we apply the empirical Huber loss exclusively to the target data to ensure that $\hat{w}_{\Lambda_{h}}$ is aligned with the target by adjusting the contrast as follows:
\begin{eqnarray}
\hat{\delta}_{\Lambda_{h}}=\arg\min_{\delta\in {R^p}}\left\{\frac{1}{n_{0}}\sum_{s\in \Lambda_{h}}\left(\sum^{n_{0}}_{i=1}\ell(y_i-z^T_i(\hat{w}_{\Lambda_{h}}+\delta))\right)+\lambda_{\delta}P_{\alpha}(\delta)\right\},
\end{eqnarray}
where $\lambda_{\delta}$ is the tuning parameter. The tuning parameter is selected by using $5$ folds cross validation. Finally, the estimation of coefficient vector $\beta$ of Huber regression model is $\hat\beta_{Oracle}=\hat{w}_{\Lambda_{h}}+\hat{\delta}_{\Lambda_{h}}$.

\begin{center}
\begin{figure}
\resizebox{\textwidth}{!}{\includegraphics[angle=0]{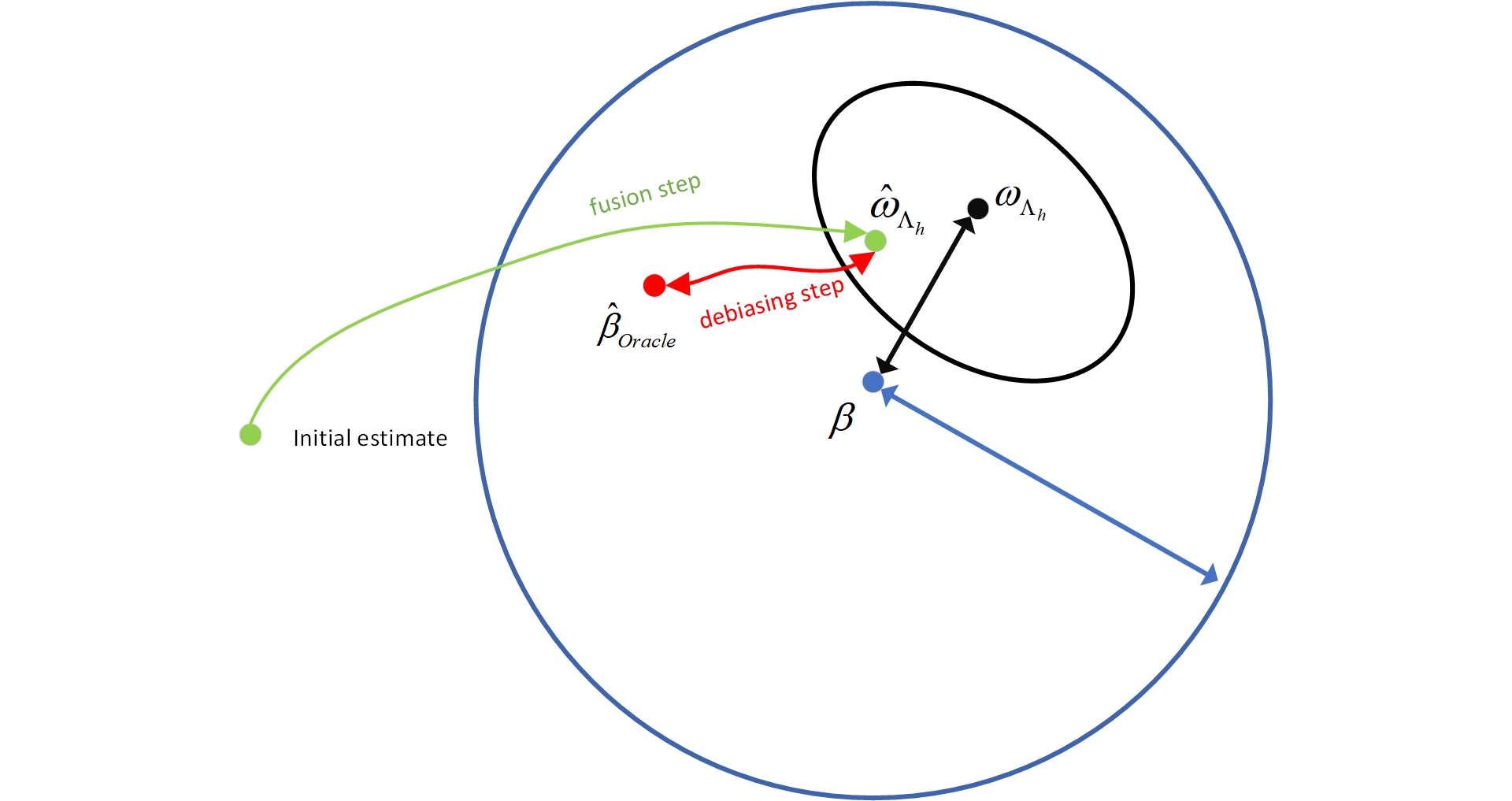}} 
\caption{A schematic of Transfer-HR.}
\end{figure}
\end{center}

\begin{table}[htbp]
\begin{tabular}{p{1.0\linewidth}}
\hline
Algorithm 1: Oracle Algorithm   \\
\hline
Input: Target data $(Z^{(0)},Y^{(0)})$ and known transferable source data $(Z^{(k)},Y^{(k)}),k\in {\Lambda_{h}}$, penalty parameters $\lambda_{w}$ and $\lambda_{\delta}$.\\
1: The fusion step:	      \\
\begin{eqnarray}
\hat{w}_{\Lambda_{h}}=\arg\min_{w\in {R^p}}\left\{\frac{1}{n_{\Lambda_{h}}+n_{0}}\sum_{s\in \Lambda_{h}}\left(\sum^{n_{s}+n_{0}}_{i=1}\ell(y_i-z^T_iw)\right)+\lambda_{w}P_{\alpha}(w)\right\},
\end{eqnarray}\\
2: The debiasing step:\\
\begin{eqnarray}
\hat{\delta}_{\Lambda_{h}}=\arg\min_{\delta\in {R^p}}\left\{\frac{1}{n_{0}}\sum_{s\in \Lambda_{h}}\left(\sum^{n_{0}}_{i=1}\ell(y_i-z^T_i(\hat{w}_{\Lambda_{h}}+\delta))\right)+\lambda_{\delta}P_{\alpha}(\delta)\right\},
\end{eqnarray}\\
Output: $\hat\beta_{Oracle}=\hat{w}_{\Lambda_{h}}+\hat{\delta}_{\Lambda_{h}}$\\
\hline
\end{tabular}
\end{table}

\subsubsection{Transferable source detection}
\def\theequation{\thesection.\arabic{equation}}

In Algorithm 1, we assume that the transferable source data set is known, which might be not realized in practice. This limitation motivates the proposal of an efficient algorithm to detect $\Lambda_{h}$. Identifying the transferable source datasets is crucial because negative transfer may occur if these datasets cannot enhance the estimation performance of the target model. Therefore, finding the transferable source data set becomes a key aspect in this section. Figure 2 shows a schematic of transferable source detection.

\begin{center}
\begin{figure}
\resizebox{\textwidth}{!}{\includegraphics[angle=0]{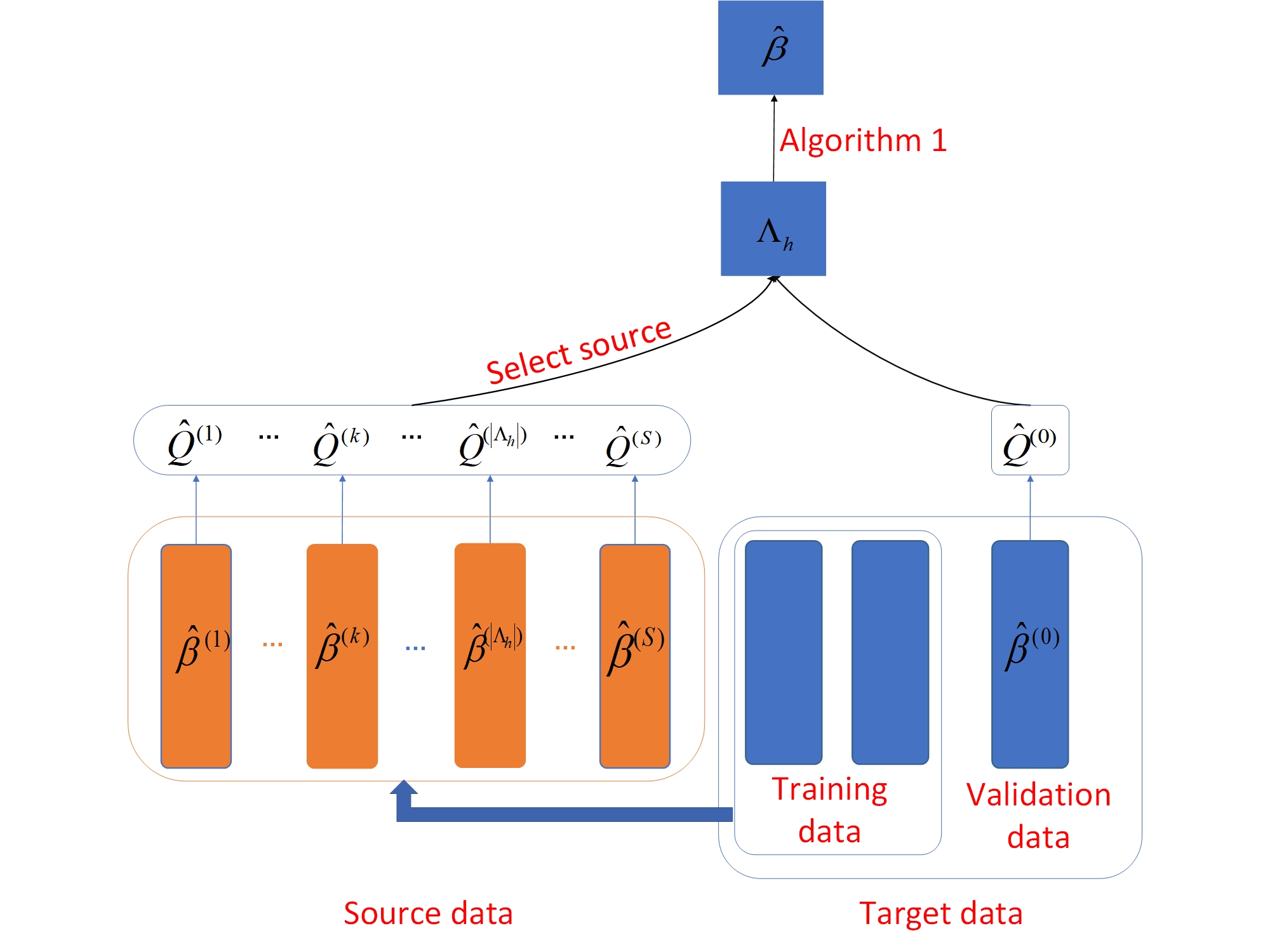}} 
\caption{A schematic of transferable source detection.}
\end{figure}
\end{center}

Motivated by Li and Song \cite{Li2024}, this section introduce an effective transfer learning detection algorithm premised on the high-dimensional Huber regression model, called Detect. The algorithm is versatile and can handle various real-world application scenarios. The main idea as follows. Firstly, the target dataset is divided into three equal folds, with two folds used as the training set and one fold as the validation set. The source datasets are then combined with the training set to estimate coefficients, which are subsequently used in conjunction with the validation set to compute the Huber loss. The average loss obtained after repeating this process three times serves as the evaluation standard. Ultimately, the source datasets meeting the specified criteria are chosen as the transferable source datasets. Please refer to Algorithm 2 for specific details.

\begin{table}[htbp]
\renewcommand{\arraystretch}{1.3} 
\begin{tabular}{p{1.0\linewidth}}
\hline
Algorithm 2: Transferable Source Detection ($\Lambda_{h}$ is unknown)   \\
\hline
Input: Target data $(Z^{(0)},Y^{(0)})$ and all source data $(Z^{(k)},Y^{(k)}),k\in \{1,\ldots,S\}$, a constant\\
  \hspace{3em} $\varepsilon_{0}>0$, penalty parameters $\{\{\lambda^{(k)}\}^{S}_{k=0}\}^{3}_{r=1}$, where $r$ is the folding number.\\
Output: $\hat\beta$ and $\hat\Lambda_{h}$.\\
1: The target data $((Z^{(0)}, Y^{(0)}))$ is randomly divided into three datasets of equal size, that is the training target data $\{(\bar{Z}^{(0)[i]},\bar{Y}^{(0)[i]})\}^{3}_{i=1}$ and the remaining data is the validation target data ${(Z^{(0)},Y^{(0)})}/\{(\bar{Z}^{(0)[i]},\bar{Y}^{(0)[i]})\}^{3}_{i=1}$. \\
2: for r=1 to 3 do\\
  \hspace{1em} $\hat\beta^{(0)}_{r}$ ${\leftarrow}$ perform the Huber regression under elastic-net penalty on the training target\\
  \hspace{2em} data, and the regularization parameter was $\lambda^{(0)[r]}$.\\
  \hspace{1em} $\hat\beta^{(k)}_{r}$ ${\leftarrow}$ run Step 1 in Algorithm 1 with the training target data, each of the source\\
  \hspace{2em} data sets $\{(Z^{(k)},Y^{(k)})\}$ , and the regularization parameter was $\lambda^{(k)[r]}$ for\\
  \hspace{2em} $k\in \{1,\ldots,S\}$.\\
  \hspace{1em} Calculate the Huber regression loss $\hat{Q}^{[r]}(\hat\beta^{(0)})$ and $\hat{Q}^{[r]}(\hat\beta^{(k)})$ on the validation target\\
  \hspace{2em} data.\\
3: $\hat{Q}^{(0)}$ ${\leftarrow}$ $\sum^{3}_{r=1}\hat{Q}^{[r]}(\hat\beta^{(0)})$, $\hat{Q}^{(k)}$ ${\leftarrow}$ $\sum^{3}_{r=1}\hat{Q}^{[r]}(\hat\beta^{(k)})$.\\
4: $\hat\Lambda_{h}$ \hspace{0.2em}  ${\leftarrow}$ $\{k\neq0:\hat{Q}^{(k)}\leq(1+\varepsilon_{0})\hat{Q}^{(0)}\}$.\\
5: $\hat\beta$ \hspace{0.7em} ${\leftarrow}$ run Algorithm 1 using $\{(Z^{(k)},Y^{(k)})\}_{k\in {\{0\}\cup\hat\Lambda_{h}}}$.\\
\hline
\end{tabular}
\end{table}

\section{Simulation experiment}
\setcounter {equation}{0}
\def\theequation{\thesection.\arabic{equation}}
In this section, we firstly consider scenarios with homogenous data and known transferable source datasets to prove the validity of the proposed method based on Huber regression model. We conduct elastic-net-penalized Huber regression with only target data (Target) ,
elastic-net-penalized Huber regression of transfer learning proposed (Oracle). Secondly, we extend our analysis to scenarios featuring heterogeneous data, where the transferable source datasets are unknown. Besides the aforementioned estimators, we perform simulation studies to compare additional approaches: (1) Naive, which naively assumes all the sources as informative without employing any detection procedure within elastic-net-penalized Huber regression; (2) Detect, which is the proposed approach that conducts informative set detection without relying on any prior knowledge in elastic-net-penalized Huber regression. We set $\alpha\in \{0.5, 1\}$ in penalized term.

\subsection{Transferable source data set is known}
\def\theequation{\thesection.\arabic{equation}}

Set $p=500$, $n_{0}=30$, $n_{k}=20, k\in \{1,\ldots,S\}$ with $S=25$. For $k\in \{0,\ldots,S\}$, the covariate vectors $x^{(k)}\in {R^{p}}$ are independent and identity distribution from the normal distribution with mean $0$ and covariance $\Sigma=[\Sigma_{jj^{'}}]_{p\times p}$. Note that the variance of the target data and the source data is the same. For target data, we set $\ell=14$, $\beta^{(0)}=(0.31_{\ell},0_{p+1-\ell})$. For $k\in {\Lambda_{h}}$, we need to satisfy
\begin{eqnarray}
\beta^{(k)}_{j}=\beta^{(0)}_{j}+\epsilon_{j}I(j\in {R_{s}}),
\end{eqnarray}
where $R_{s}=1,\ldots,100$, $\epsilon_{j}\sim{N(0,h^2/10000)}$. We have that $E||\beta^{(k)}-\beta^{(0)}||_{1}=h$, where $h$ represents a measure of similarity, taking values from the set $\{4,6,8,10\}$. We also consider different numbers of transferable source datasets denoted by $|\Lambda_{h}|$, which can come from the set $\{0,5,10,15,20\}$.

To evaluate the performance of various methods, we compute the mean square error between the parameter estimator $\hat\beta$ and the true value of the parameter $\beta^{(0)}$, that is $||\hat\beta-\beta^{(0)}||^2_{2}$. Each setting is replicated by 200 times and the results of each experiment is averaged to get the result value of each point in Figures 3-8.

As shown in Figures 3-8, when $h$ is held constant within the same distribution, a greater number of transferable sources provide more effective data for the target, thereby reducing estimation errors in the target model. Similarly, when the number of transferable source datasets is the same, smaller values of $h$ lead to smaller Oracle estimation errors. This indicates that the similarity between the target data and the source data contributes to enhancing the efficiency of transfer learning. Furthermore, the estimated performance of Oracle algorithm is better than that of Target algorithm, which shows that it has a positive effect in the Target studies without negative transfer.

\begin{center}
\begin{figure}
\resizebox{\textwidth}{!}{\includegraphics[angle=0]{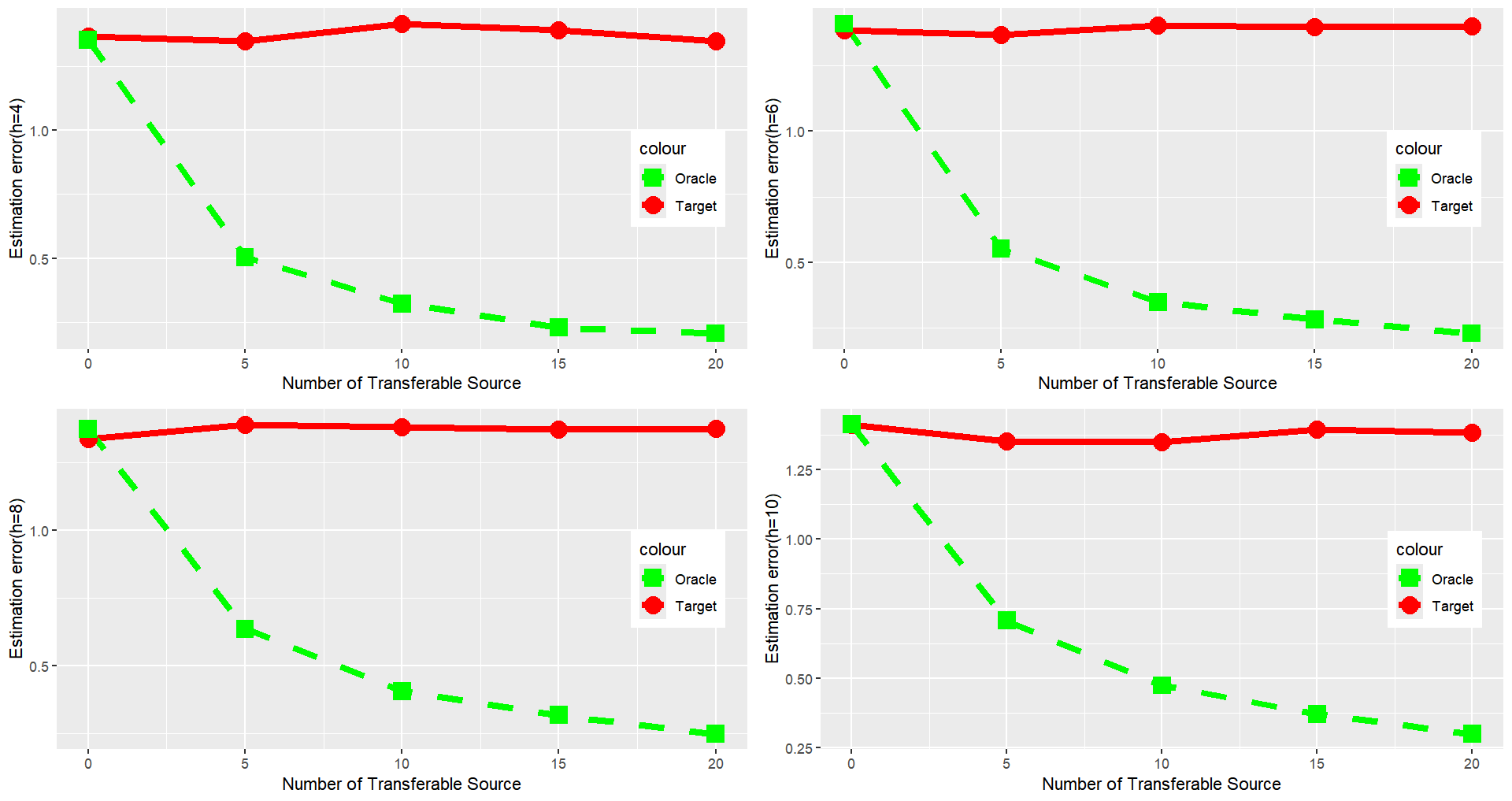}} 
\caption{Mean square error of estimation under the condition of similarity degree of $h = 4, 6, 8, 10$ and $\alpha=1$ for normal error distribution.}
\label{Normal}
\end{figure}
\end{center}

\begin{center}
\begin{figure}
\resizebox{\textwidth}{!}{\includegraphics[angle=0]{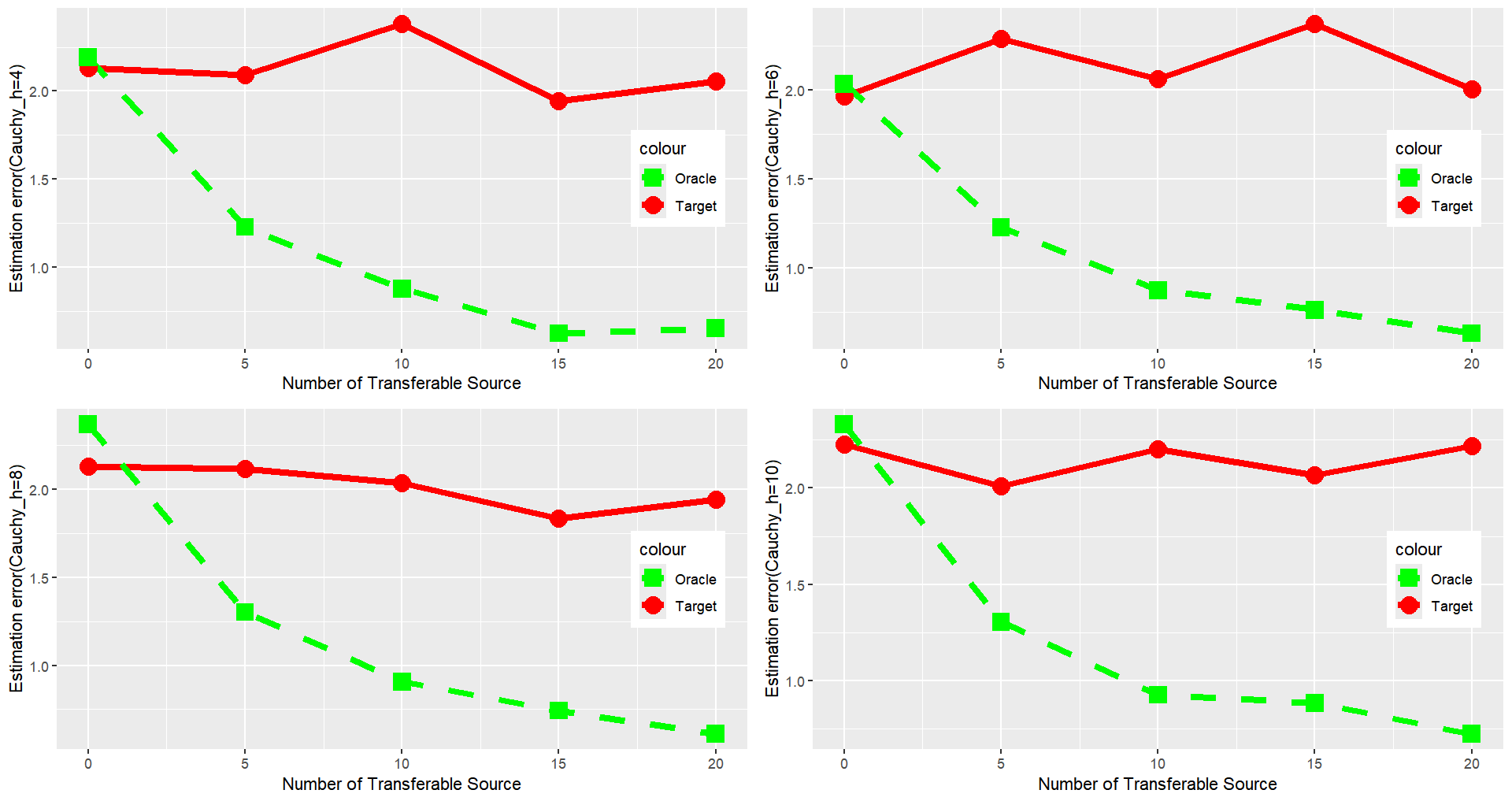}} 
\caption{Mean square error of estimation under the condition of similarity degree of $h = 4, 6, 8, 10$ and $\alpha=1$ for cauchy error distribution.}
\label{Cauchy}
\end{figure}
\end{center}

\begin{center}
\begin{figure}
\resizebox{\textwidth}{!}{\includegraphics[angle=0]{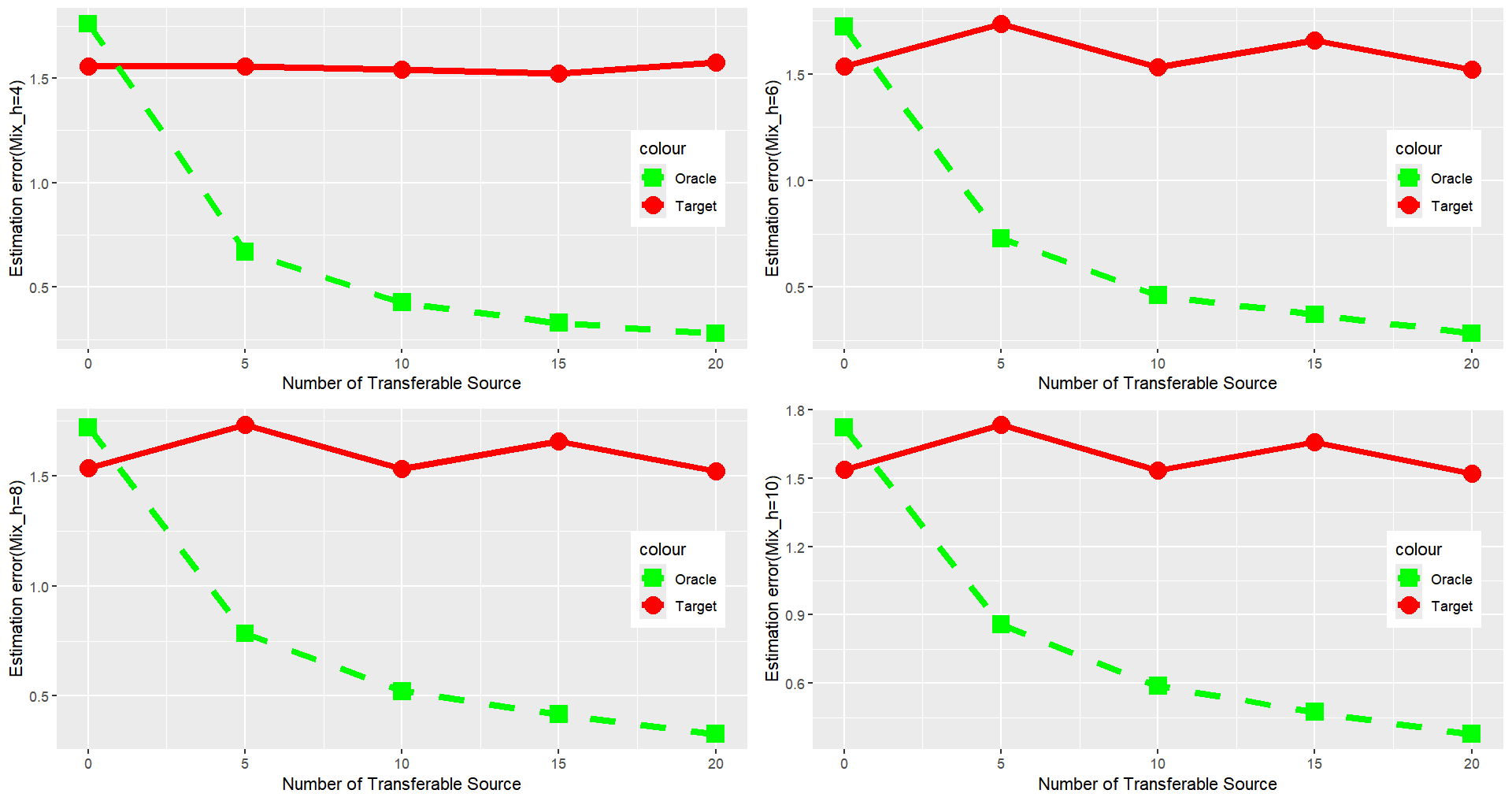}} 
\caption{Mean square error of estimation under the condition of similarity degree of $h = 4, 6, 8, 10$ and $\alpha=1$ for mixed normal error distribution.}
\label{Mix}
\end{figure}
\end{center}
\begin{center}
\begin{figure}
\resizebox{\textwidth}{!}{\includegraphics[angle=0]{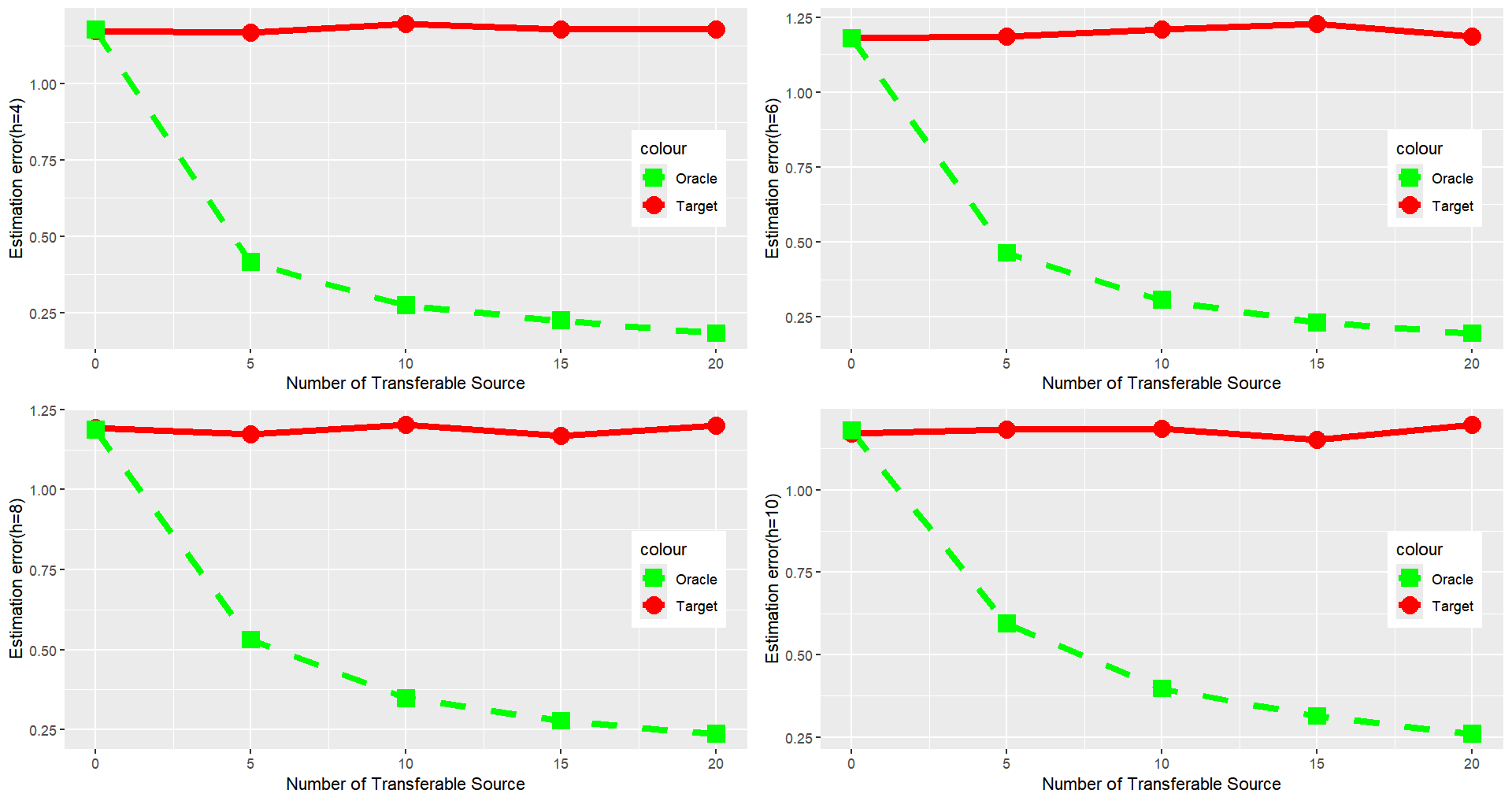}} 
\caption{Mean square error of estimation under the condition of similarity degree of $h = 4, 6, 8, 10$ and $\alpha=0.5$ for normal error distribution.}
\label{Norm}
\end{figure}
\end{center}

\begin{center}
\begin{figure}
\resizebox{\textwidth}{!}{\includegraphics[angle=0]{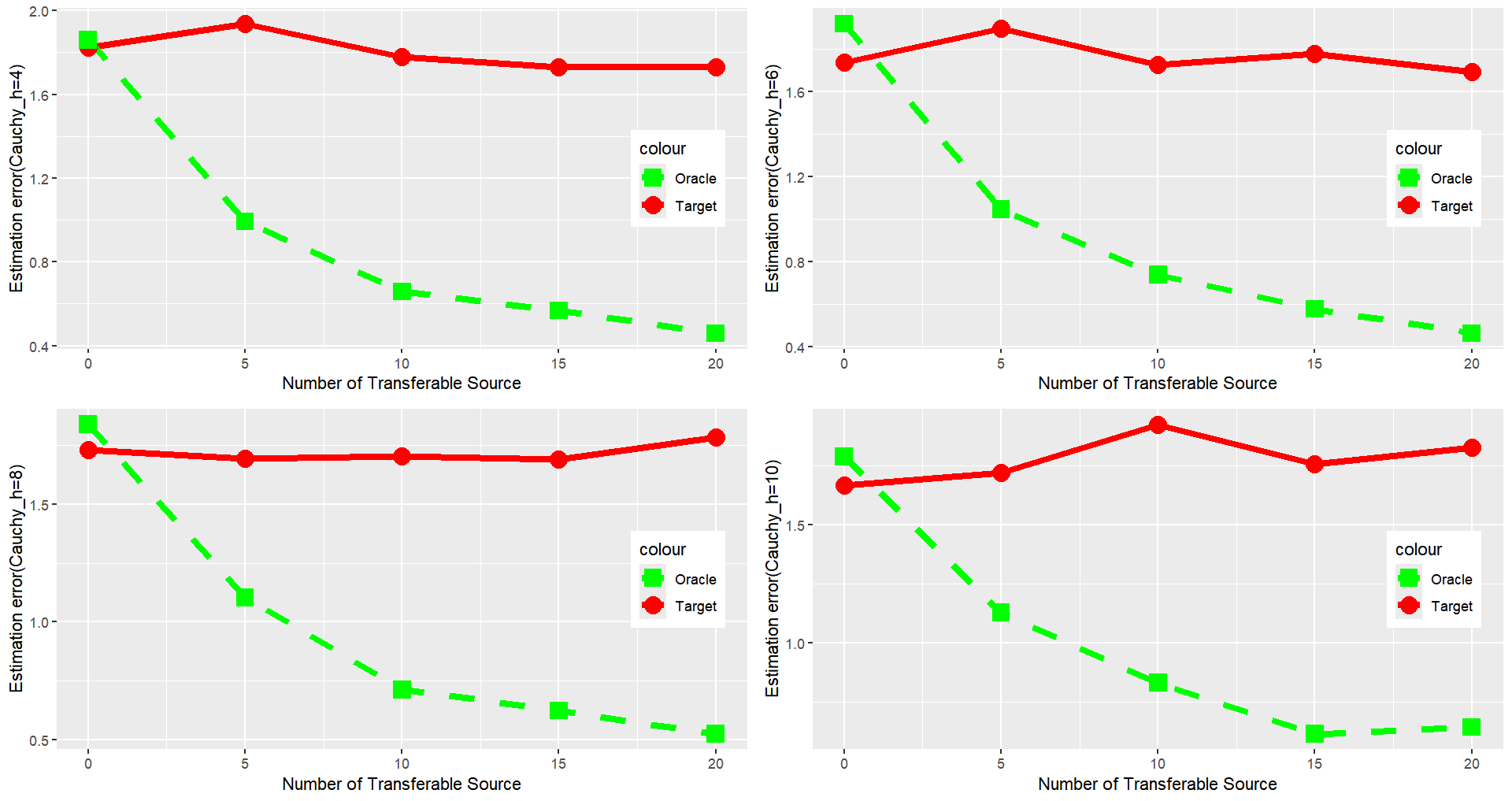}} 
\caption{Mean square error of estimation under the condition of similarity degree of $h = 4, 6, 8, 10$ and $\alpha=0.5$ for cauchy error distribution.}
\label{Cauchy}
\end{figure}
\end{center}

\begin{center}
\begin{figure}
\resizebox{\textwidth}{!}{\includegraphics[angle=0]{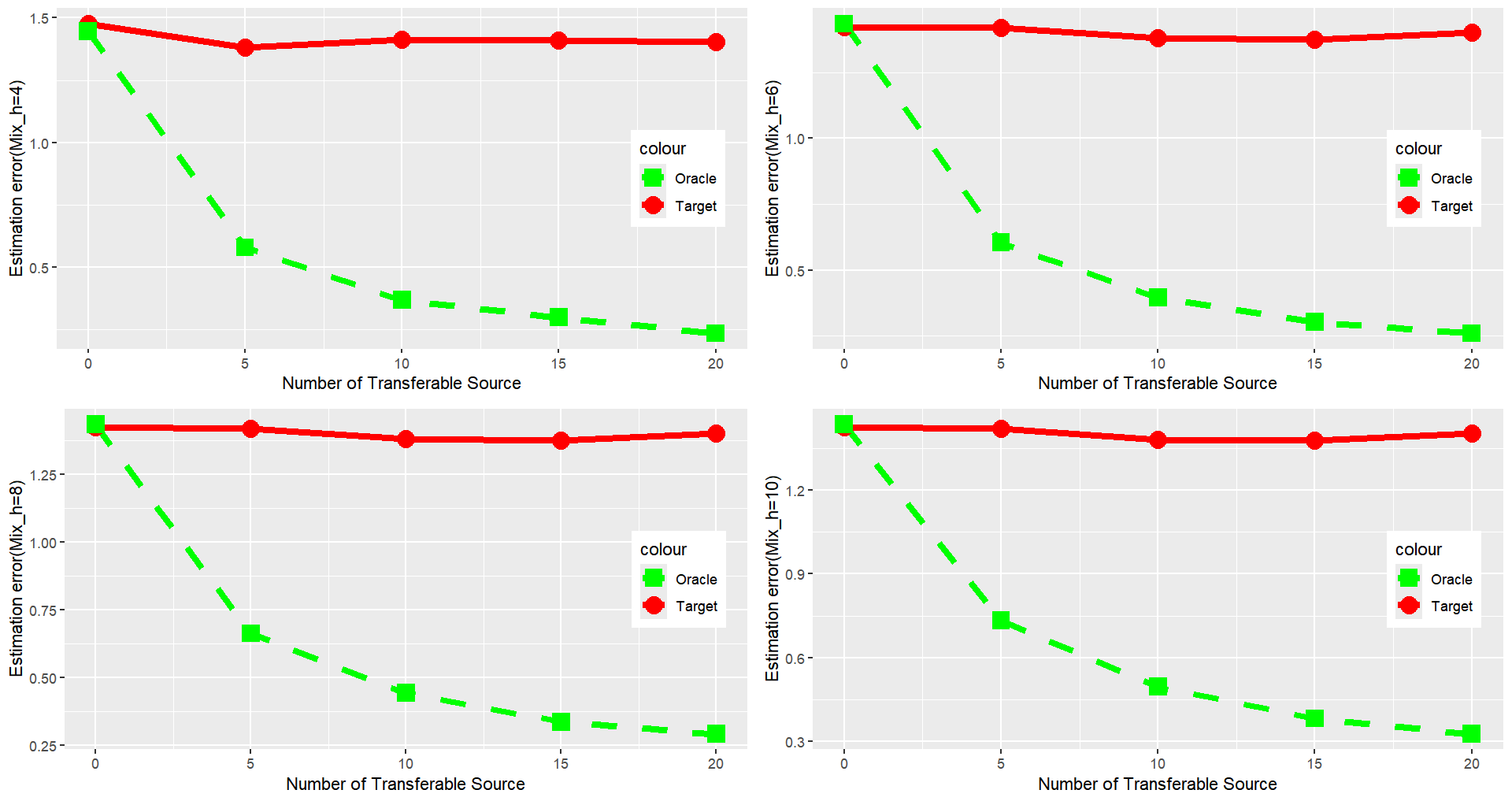}} 
\caption{Mean square error of estimation under the condition of similarity degree of $h = 4, 6, 8, 10$ and $\alpha=0.5$ for mixed normal error distribution.}
\label{Mix}
\end{figure}
\end{center}

\subsection{Transferable source data set is unknown}
\def\theequation{\thesection.\arabic{equation}}

We investigate the performance of the transferable source detection algorithm in this section. Set $p=500$, $n_{0}=100$, $n_{k}=100, k\in \{1,\ldots,S\}$ with $S=10$. For target dataset, $x^{(0)}\in {R^{p}}$ are independent and identity distribution from the normal distribution with mean $0$ and covariance $\Sigma$, where $\Sigma$ is a $p$ dimensional identity matrix. For each source dataset, $x^{(k)}\sim N(0, \Sigma^{(k)})$, where $\Sigma^{(k)}$ exhibits a Toeplitz structure
\begin{eqnarray*}
	\Sigma^{(k)}=
\left(
\begin{array}
	{cccccccc}
	\sigma_{0} &\sigma_{1} &\sigma_{2} &$\ldots$ &\sigma_{p-1}\\
	\sigma_{1} &\sigma_{0} &\sigma_{1} &$\ldots$ &\sigma_{p-2}\\
	\sigma_{2} &\sigma_{1} &\sigma_{0} &$\ldots$ &\sigma_{p-3}\\
    $\vdots$   &$\vdots$    &$\vdots$   &$\vdots$  &$\vdots$\\
    \sigma_{p-1} &\sigma_{p-2} &$\ldots$ &\sigma_{1} &\sigma_{0}\\
	\end{array}
\right)
\end{eqnarray*}
with its first row given as
\begin{eqnarray*}
	(\sigma_{0},\ldots,\sigma_{p-1})=(1,\underbrace{1/(s+1),\ldots,1/(s+1)}_{2s-1},0_{p-2s}),
\end{eqnarray*}
for all $i=1,\ldots,n_{k}, k=1,\ldots,S$.\\
\\
$\bullet$	if $k\in{\Lambda_{h}}$, let\\
\begin{eqnarray}
	\beta^{(k)} = \beta^{(0)}+(h/p)R^{(k)}_{p}
\end{eqnarray}
where $R^{(k)}_{p}$ represents $p+1$ independent Rademacher vectors.\\
$\bullet$	if $k\notin{\Lambda_{h}}$, let\\
\begin{equation}\label{eq2}
    \beta_{j}^{(k)}=\left\{
    \begin{aligned}
        &0.5+2he^{(k)}_{j}, && j\in\{j+1,\ldots,2l\}\cup{M^{(k)}}\\
        &2he^{(k)}_{j}, && \text{otherwise}
    \end{aligned}
    \right.
\end{equation}
\\
where $\beta_{j}^{(k)}$ is the $j-th$ element of $\beta^{(k)}$, $M^{(k)}$ is the randomly generated set of size $l$ from
the set $\{2l + 1, ... , p+1\}$, and $e^{(k)}_{j}$ is the Rademacher variable.
\\

We consider $h\in\{30,60\}$ and different numbers of transferable source datasets $|\Lambda_{h}|\in\{0,2,4,6,8,10\}$. Furthermore, to understand whether the proposed method can solve the heavy tail problem. For error distribution $\epsilon^{(k)}_{i} (i=1,\ldots,n_{k},\ k=0,\ldots,S)$, we set $\epsilon^{(k)}_{i}\sim{N(0,1)}$,\ $\epsilon^{(k)}_{i}\sim{Cauchy(0,1)}$ and\ $\epsilon^{(k)}_{i}\sim{0.9N(0,1)+0.1N(0,100)}$ under homogeneous and heterogeneous design. We conduct 200 independent simulation experiments and show these results, which include Target, Oracle, Naive and Detect in Figures 9-14.

Figures 9-14 depict the coefficient estimation errors of different methods for normal distribution, cauchy distribution, and mixed normal distribution. As $|\Lambda_{h}|$ increases, the Oracle algorithm exhibits superior estimation accuracy compared to the Target algorithm. This trend highlights the superiority of Oracle in these scenarios. Initially, when $|\Lambda_{h}|$ is small, the Naive method exhibits higher estimation errors than the Target method, suggesting a potential negative transfer phenomenon. However, as $|\Lambda_{h}|$ grows, the estimation error of Naive gradually decreases beyond that of Target, indicating improved performance due to augmented effective information derived from the target dataset. Overall, transfer learning methods demonstrate advantages over the traditional Target method in enhancing estimation performance.

In addition, When $|\Lambda_{h}|=0$, the estimation error of the Detect method surpasses that of the Target method, as Detect relies solely on target data for estimation. As $|\Lambda_{h}|$ increases, the estimation error of Detect gradually diminishes, indicating improved performance over time. No matter what the error distribution, the estimation error of Detect falls between that of the Naive and Oracle methods. Moreover, as $h$ increases, the errors of different methods also increase because the amount of information between the source data and the target data has decreased.

By comparing Figures 9-14, the Oracle and Detect methods outperform the Target estimation method for various error distribution. This superiority across different error distributions validates the efficacy of the transfer learning approach, particularly when employing the Huber regression model, which effectively addresses issues of data heterogeneity. Additionally, Detect algorithm demonstrates its robustness across different values of $h$. This robustness indicates that it can adapt to variations between source and target data distributions, thereby enhancing its reliability in practical applications. In summary, these findings not only underscore the comparative advantages of the Oracle and Detect methods over the Target method but also confirm the robustness and adaptability of the Detect method in the context of transfer learning and heterogeneous data environments.

\begin{center}
\begin{figure}
\resizebox{\textwidth}{!}{\includegraphics[angle=0]{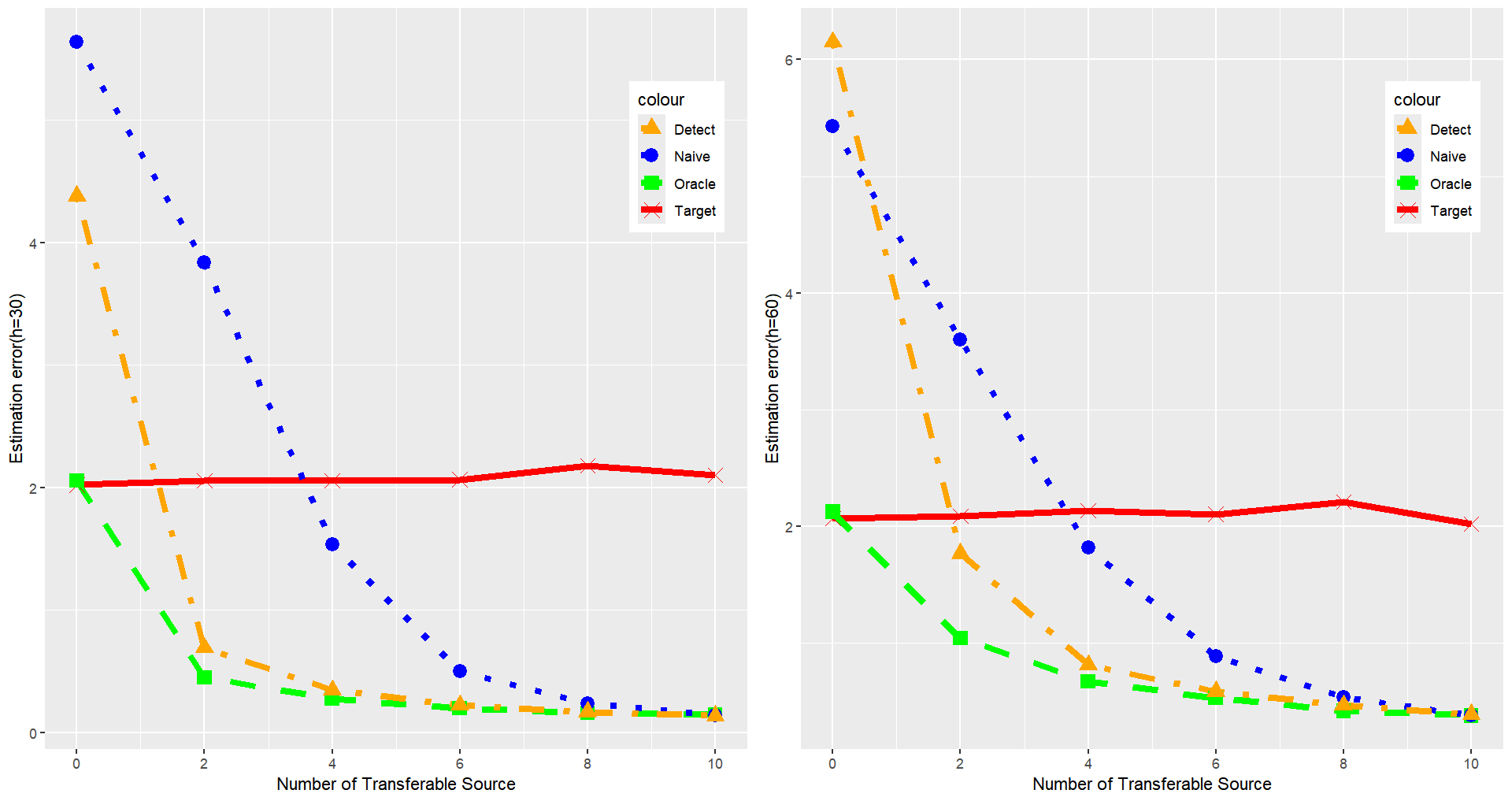}} 
\caption{Coefficient estimation error on $\hat\beta$ among different $\Lambda_{h}$ with $h = 30, 60$ and $\alpha=1$ for normal error distribution.}
\label{Normal}
\end{figure}
\end{center}

\begin{center}
\begin{figure}
\resizebox{\textwidth}{!}{\includegraphics[angle=0]{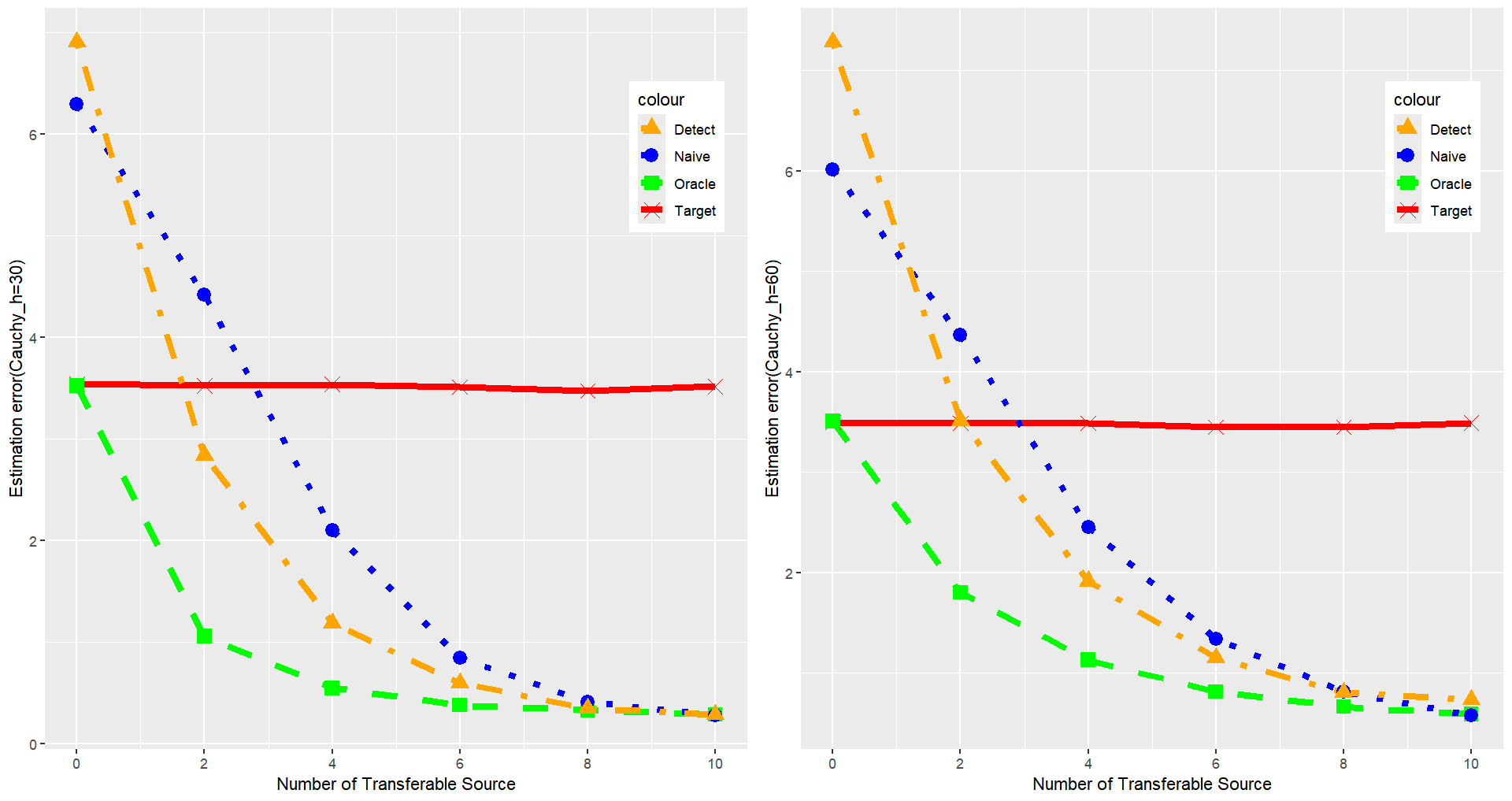}} 
\caption{Coefficient estimation error on $\hat\beta$ among different $\Lambda_{h}$ with $h = 30, 60$ and $\alpha=1$ for cauchy error distribution.}
\label{Cauchy}
\end{figure}
\end{center}

\begin{center}
\begin{figure}
\resizebox{\textwidth}{!}{\includegraphics[angle=0]{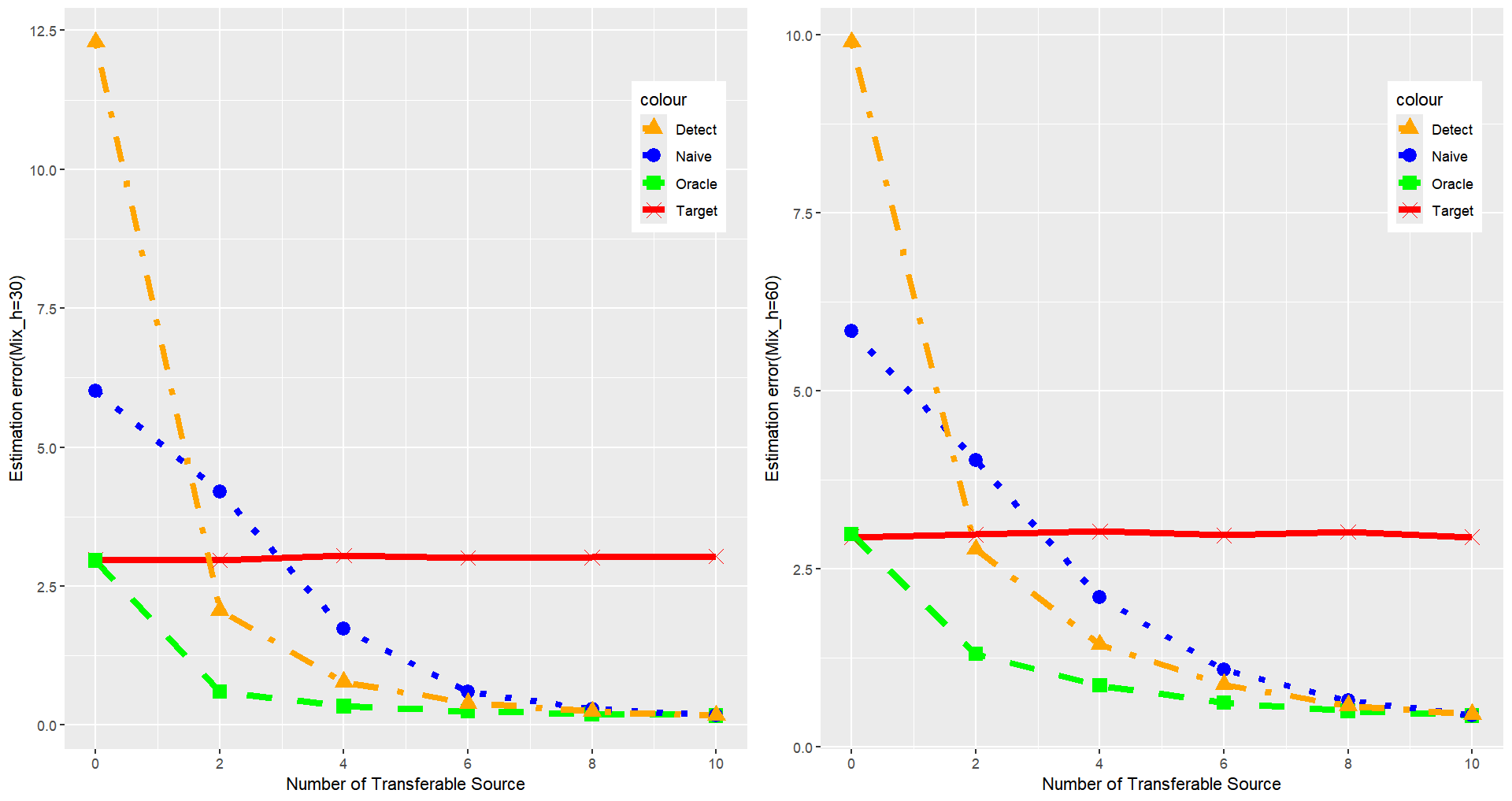}} 
\caption{Coefficient estimation error on $\hat\beta$ among different $\Lambda_{h}$ with $h = 30, 60$ and $\alpha=1$ for mixed normal error distribution.}
\label{Mix}
\end{figure}
\end{center}

\begin{center}
\begin{figure}
\resizebox{\textwidth}{!}{\includegraphics[angle=0]{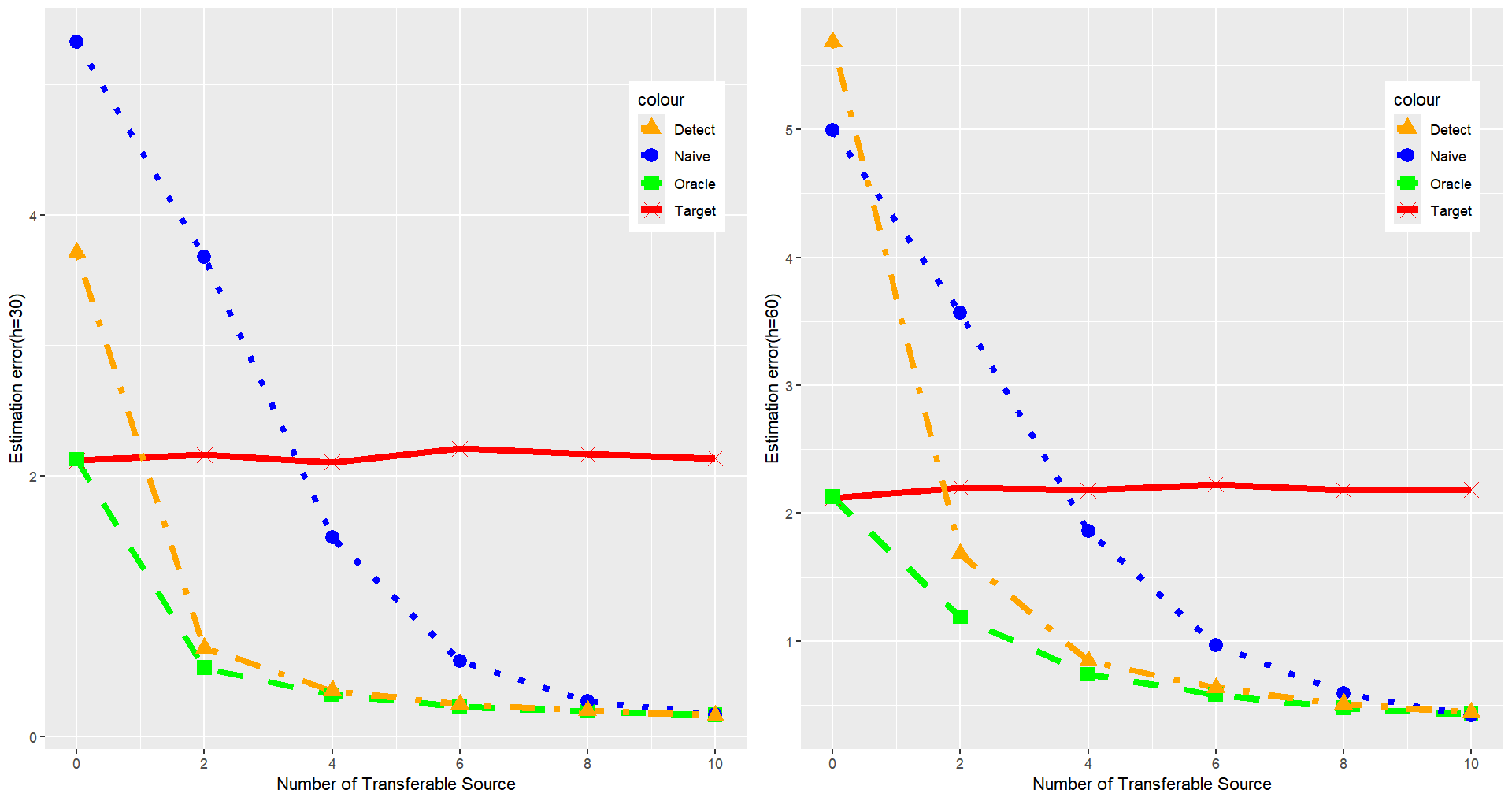}} 
\caption{Coefficient estimation error on $\hat\beta$ among different $\Lambda_{h}$ with $h = 30, 60$ and $\alpha=0.5$
for normal error distribution.}
\label{Normal}
\end{figure}
\end{center}

\begin{center}
\begin{figure}
\resizebox{\textwidth}{!}{\includegraphics[angle=0]{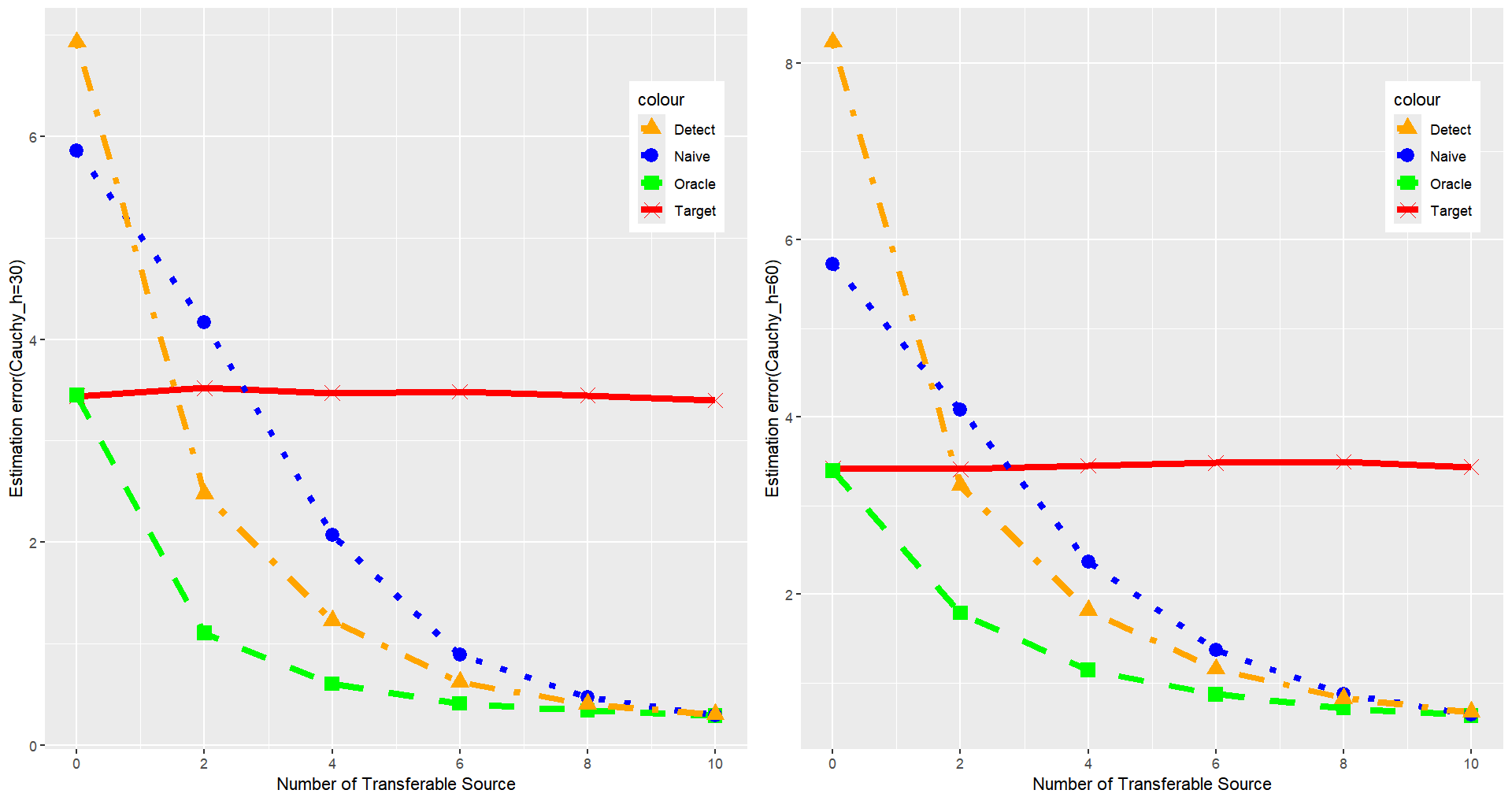}} 
\caption{Coefficient estimation error on $\hat\beta$ among different $\Lambda_{h}$ with $h = 30, 60$ and $\alpha=0.5$
for cauchy error distribution.}
\label{Cauchy}
\end{figure}
\end{center}

\begin{center}
\begin{figure}
\resizebox{\textwidth}{!}{\includegraphics[angle=0]{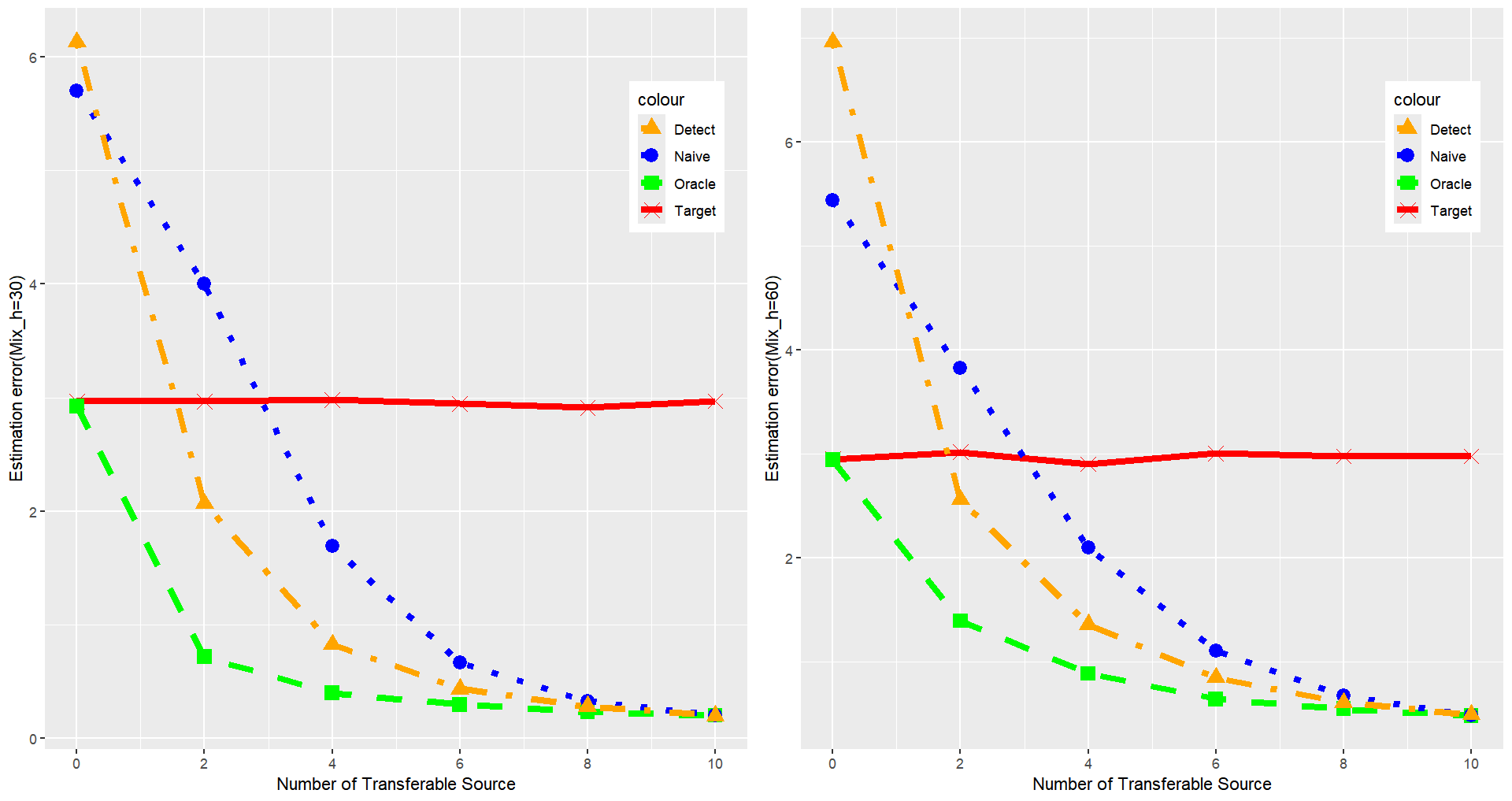}} 
\caption{Coefficient estimation error on $\hat\beta$ among different $\Lambda_{h}$ with $h = 30, 60$ and $\alpha=0.5$
for mixed normal error distribution.}
\label{Mix}
\end{figure}
\end{center}

\section{Applications to real data}
\setcounter {equation}{0}
\def\theequation{\thesection.\arabic{equation}}

In this study, we utilized data from Japan's National Institute for Materials Science (NIMS) Superconducting Material Database, encompassing 21,263 superconductors characterized by 82 variables. This dataset was previously processed by Hamidieh \cite{Hamidieh2018}, focusing on statistical modeling aimed at predicting the critical temperature of these superconductor materials. Superconductors hold immense practical value, ranging from advanced medical imaging applications like Magnetic Resonance Imaging (MRI) to the creation of high magnetic fields in facilities such as the Large Hadron Collider at CERN. Moreover, the potential for superconductors to revolutionize the energy sector lies in their capability to enable frictionless transmission of electricity, promising significant reductions in energy loss during distribution and utilization.

Before conducting the analysis, we conducted several preprocessing steps on the raw data. This involved standardizing continuous variables and applying one-hot encoding to categorical variables. Subsequently, we compiled a finalized dataset comprising 87 covariates and 1 response variable known as the critical temperature. This paper uses these covariate variables and dependent variable to structure the Huber regression model. This model has an important effect on the critical temperature of superconductor material.

The Number of elements in the superconducting material has a complex effect on superconductivity, which is influenced by the behaviour and interactions of electrons in the material, which in turn are closely related to the element composition of the material. Therefore, we consider the superconducting material samples with four elements as the target data, and the remaining samples as the source data. To evaluate the performance of the transfer learning method, we randomly select  80 percent of data as the target training data and the remaining data as the target test data.

Based on above standard, the samples are divided into 8 source datasets and 1 target dataset in Table 1. We apply a transferable detection algorithm to identify suitable source datasets. Algorithm 2 identifies source datasets 7 and 8 as effective transferable source datasets. Subsequently, Algorithm 1 is employed to estimate parameters for the Huber regression model, and the mean square error of the target test set is computed. Figure 15 presents these results. Through comparing the performance of Target, Naive and Detect, it is found that the effective transferable source datasets selected by algorithm 2 can assist in constructing the target model, demonstrating a positive transfer effect that improves the target task.

\newpage
\begin{table*}
\centering
\caption{Sample sizes.}\label{tab:parametervalues}
\begin{tabular*}{\hsize}{@{}@{\extracolsep{\fill}}ccccccccccc@{}}
\toprule
$Data set$  & 1   & 2    & 3    & 4    & 5    & 6   & 7  & 8   & $Training$  & $Test$ \\
\midrule
$sample$    & 285 & 3280 & 3895 & 5792 & 2666 & 774 & 61 & 14  & 3597        & 899 \\
\bottomrule
\end{tabular*}
\end{table*}

\begin{center}
\begin{figure}
\resizebox{\textwidth}{!}{\includegraphics[angle=0]{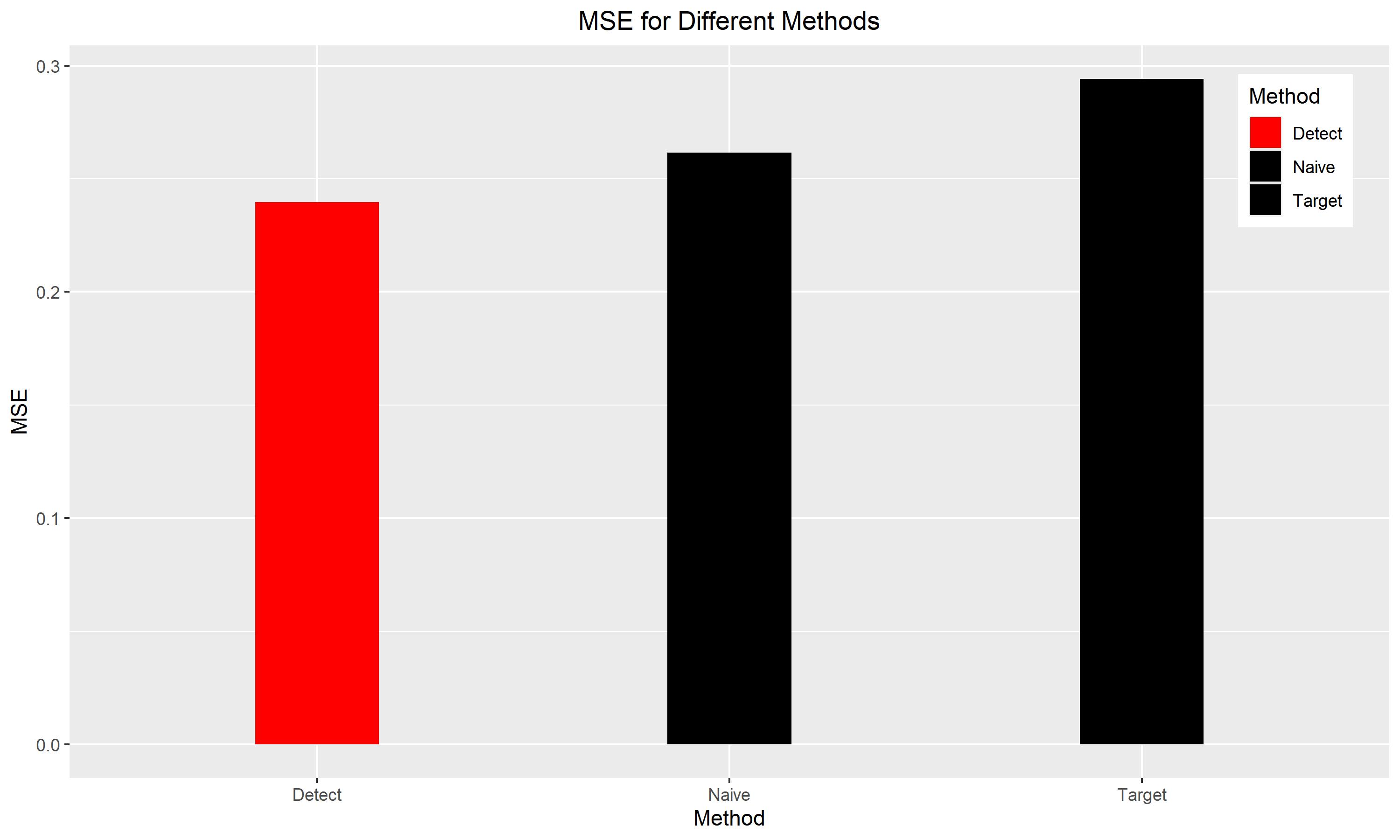}} 
\caption{Mean square error of different algorithms.}
\label{Mix}
\end{figure}
\end{center}

\section{Discussion}
\setcounter {equation}{0}
\def\theequation{\thesection.\arabic{equation}}

This paper proposes a transfer learning algorithm for robust regression models in high-dimensional data, specifically designed for scenarios where the transferable source data set is known. It tackles heteroscedasticity issues inherent in such data and achieves robust estimation. Furthermore, in cases where the transferable source data is unknown, an effective source detection algorithm is developed to achieve the purpose of transferable source identification. The algorithm's feasibility is demonstrated through numerical simulations and empirical studies, although it currently lacks theoretical proof of its effectiveness. Nowadays, most papers utilize the $\ell_{1}$ norm to measure the similarity between the source models and target model. Subsequently, researchers could consider measuring their distance using trigonometric functions such as sine and cosine values.

\section*{Acknowledgements}


\makeatletter
\renewenvironment{thebibliography}[1]
{\section*{\refname}%
\@mkboth{\MakeUppercase\refname}{\MakeUppercase\refname}%
\list{\@biblabel{\@arabic\c@enumiv}}%
{\settowidth\labelwidth{\@biblabel{#1}}%
\leftmargin\labelwidth \advance\leftmargin\labelsep
\advance\leftmargin by 2em%
\itemindent -2em%
\@openbib@code
\usecounter{enumiv}%
\let\p@enumiv\@empty
\renewcommand\theenumiv{\@arabic\c@enumiv}}%
\sloppy \clubpenalty4000 \@clubpenalty \clubpenalty
\widowpenalty4000%
\sfcode`\.\@m} {\def\@noitemerr
{\@latex@warning{Empty `thebibliography' environment}}%
\endlist}
\renewcommand\@biblabel[1]{}
\makeatother


\end{document}